# Natural Interaction Modalities for Human-CPS Interaction in Construction Progress Monitoring

Srijeet Halder[1], Kereshmeh Afsari[2], and Alireza Shojaei[3]

[1]Assistant Professor, Department of Sustainable Technology and the Built Environment, Appalachian State University, Boone, NC, United States. Email: halders1@appstate.edu. Corresponding author.

[2]Assistant Professor, Myers Lawson School of Construction, Virginia Tech, Blacksburg, VA, United States. Email: keresh@vt.edu.

[3]Assistant Professor, Myers Lawson School of Construction, Virginia Tech, Blacksburg, VA, United States. Email: shojaei@vt.edu.

## ABSTRACT

This article explores natural interaction modalities for human-cyber-physical systems (CPS) interaction in construction. CPS has been applied in construction for many purposes with the promise of improving the safety and productivity of construction operations. However, there is little research on human-CPS interaction in construction. This study proposes two methodologies for human-CPS interactions for construction progress monitoring – a) hand gesture interaction using transfer learning, and b) voice command interaction using natural language processing. User studies with thirty-two users validated the generalizability of the proposed methodologies. The proposed hand gesture recognition method achieved higher accuracy (99.69% vs 87.72%) and speed (36.05ms vs 578.91ms) than the proposed voice command recognition method, though users performed the progress monitoring task more correctly with voice commands than hand gestures (88% vs 66.1%). The main contribution of the study is the development of an ML pipeline and computational framework to recognize hand gestures and voice commands without the need for a large training dataset for human-CPS interaction.

## INTRODUCTION

Cyber-Physical System (CPS) is an integrated network of cyber and physical components like Building Information Modelling (BIM), robots, artificial intelligence (AI), etc. Construction 4.0, which is the industry-wide push for automation in the construction industry, promotes the use of CPS in construction for many applications [1]. Use of CPS has been proposed for many applications in the architecture, engineering, construction, and operations (AECO) industry, such as progress monitoring [2], safety monitoring [3], temporary structures monitoring [4], crane operations [5], structural health monitoring [6], smart city management [7], facility management [8], etc. In a previous study, the author proposed a robotic CPS framework for construction progress monitoring [9]. Construction progress monitoring is a crucial process in construction that ensures that the construction is progressing in a manner consistent with the original plans [10]. Using robotic CPS can improve the productivity and safety of the progress monitoring process by partially automating the process with robots [2].

However, traditionally the interaction between humans and CPS are operationalized with traditional interfaces such as keyboards, mouse, and touchscreens [11], which may not be suitable for use in the construction environment. In this context, voice command interaction and hand gesture recognition have emerged as promising alternative modalities for human-CPS interaction [12,13]. Hand gesture recognition technology allows users to interact with CPS systems using natural hand movements [12]. This technology has been applied in various domains such as gaming, healthcare, and robotics, but its potential in the construction industry remains largely unexplored. Hand signaling is frequently used among construction workers to communicate from a distance [14]. Therefore, construction workers are already familiar with hand gesture-based communication, which can be leveraged to facilitate hand gesture-based interactions between humans and CPS. On the other hand, voice-based systems have seen increased attention in the past few years with the advent of many voice-based hardware and software like Google Home, Alexa, Siri, etc. [15]. These voice-based systems were found to provide an intuitive method of information access and improve productivity by past researchers [15]. Voice and hand gestures are also key components of human-to-human communication [16]. These interactions, which utilize human language and human behavior in the interaction with computer systems, are called natural interactions [17]. They are designed to reuse the existing skills of the users in terms of speech and hand movements which reduces the corresponding cognitive load of the interactions [17]. Therefore, these two interaction modalities are studied in this research.

Previous research on hand gesture-based interaction between humans and computer systems has largely used motion sensors [12]. Recently, with advancements in computer vision techniques, researchers have used vision-based hand gesture recognition, especially for human-robot collaboration [18–20]. However, these methods either used specialized equipment like Kinect or deep learning methods like Convolutional Neural Networks (CNN). Deep learning techniques like CNN require large training datasets and/or powerful computing hardware to achieve good prediction accuracy and a high prediction rate [21]. Therefore, this research presents a methodology to recognize hand gestures from RGB images for human interactions with a CPS for construction progress monitoring. To reduce the need for large training data, a transfer learning approach was used with a pre-trained machine learning (ML) model called Mediapipe Hands [22]. For voice command recognition, previous research has utilized dedicated voice assistant hardware which has limited vocabulary and requires additional hardware [15,23]. In this study, a voice recognition model was developed using Natural Language Processing (NLP) by recording and extracting voice inputs from the user and calculating similarities of the voice input with a list of possible commands for a CPS. The model was evaluated with an in-house voice dataset for validation. Both the methodologies for hand gesture and voice command recognition were evaluated through a user study with 32 user participants. For evaluation purposes, the use case of construction progress monitoring was used during the user studies, which has been considered an important application of CPS in construction [2,24].

The main objective of this article is to demonstrate the potential of hand gestures and voice commands as alternative human-CPS interaction modalities in construction progress monitoring. The novelty of the research lies in the development of a transfer-learning and NLP-based method for hand gesture and voice command recognition, without the need for an extensive training dataset. The article describes the technical details of the proposed method and presents the results of the user study. In the following section, a background literature review on CPS in construction, hand gesture recognition, and voice commands as interaction modalities is presented. Next, the overall methodology used in the study is explained in the Methodology section. Next, the proposed method of vision-based hand gesture recognition using a transfer learning approach is explained. After that, the computational framework for recognizing voice commands and using them as input for CPS is presented. Then, the results of the user study conducted to evaluate the proposed methods are presented in the User Study section. In the Discussion section, the implications of the findings of the study are discussed. Finally, the article discusses the limitations of the study and concludes with directions for future research in this area.

## BACKGROUND

### CPS in Construction

Coined in 2006, the term CPS gained recognition as a significant research domain following its identification by the US National Science Foundation (NSF) in 2008 [25]. It refers to interconnected networks of computational and physical components that possess the capability to perceive the real world and interact with it through sensors and actuators [24,26]. By autonomously adapting to changing environments and user requirements, CPS can introduce autonomous functionalities to physical processes [27]. CPS has found wide-ranging applications in industries such as automotive, manufacturing, power generation, and construction [27,28]. In the construction industry, CPS has been increasingly applied to real-time monitoring of construction progress, safety, and quality [28,29]. CPS in construction typically involves the use of sensors, data communication networks, and software algorithms to collect and analyze data from construction sites [29]. Within this context, humans play a vital role by providing feedback to CPS, enabling the handling of complex tasks within unstructured environments through the integration of human cognitive abilities and the autonomous behavior of CPS [27]. When humans are involved in the feedback loop of CPS, it is referred to as human-in-the-loop CPS or HiLCPS (sometimes abbreviated as HiTL-CPS) [3,27,30,31]. Humans may assume one of three roles in HiLCPS – a) data acquisition (collect input data), b) state inference (provide intelligence), and c) actuation (manipulate physical objects) [31]. The physical-side components are responsible for perceiving the environment and generating data, which is then processed by the cyber-component. Subsequently, the cyber-component provides feedback and control to the physical components [32].

### Hand Gesture Recognition as an Interaction Modality

Various modalities are explored for human-CPS interactions. While classical human-computer interaction methods utilized mouse and keyboard-based interaction, modern human-computer interaction methods have explored other ways of providing input to the system from humans, for example, gaze following, voice command recognition, facial expression recognition, and hand gesture recognition [33].

Hand gestures can be categorized as static or dynamic. Static gesture recognition relies on the hand's position and orientation, whereas dynamic gesture recognition tracks the hand's motion over time [19]. Ahmed et al. [12] employed wearable sensors to detect and recognize hand gestures for human-CPS interaction and created a reliable method of controlling a robotic CPS through hand gesture control. However, these sensors can be expensive, complex, and unnatural [18]. Other methods for detecting hand gestures include the use of RGB-D sensors like Kinect 1.0 [20] and time-of-flight sensors [34], both of which require specialized equipment. Therefore, Nuzzi et al. [35] employed the CNN technique to recognize hand gestures from images, utilizing the CNN's ability to extract features as well as classify them based on patterns learned from a training dataset. A previous study by Wang and Zhu [14] using CNN, specifically the ResNext-101 architecture, for hand gesture classification achieved an accuracy of 93.3% with a 170ms prediction time on average.

However, deep learning methods like CNN demand substantial training data, powerful computing hardware, and extended training time [21]. For example, Wang and Zhu [14] used a training dataset of 364 RGB-D videos for 11 gestures recorded in 7 different background settings amounting to 426,602 static frames. The transfer learning approach offers a solution by leveraging pre-trained machine learning models, originally trained for one problem, and adapting them to solve related problems [21]. Instead of training a completely blank model, transfer learning takes advantage of the knowledge acquired by an existing model on a different dataset to enhance the performance of a new model [36]. For example, Kolar et al. [21] successfully re-trained the VGG-16, a popular deep learning architecture pre-trained on the ImageNet dataset, to detect guardrails on a construction site, achieving an accuracy of 96.5% with only 4000 training images. Another approach, known as feature representation transfer, utilizes the output of a pre-trained model for one problem as input for another model addressing a different but related problem [37]. Based on the success found in this past research with transfer learning, the current study adopts a feature representation transfer learning approach for hand gesture recognition, enabling the reuse of a pre-trained model as a feature generator for another classification model.

Mediapipe Hands is a pre-trained model developed by Google Research for detecting 21 different landmarks on a human hand [22]. The Mediapipe Hands model exhibits high accuracy and speed in hand landmark detection even on low-power mobile processors [22]. The model has been trained using a combination of six thousand public images, ten thousand in-house images, and one hundred thousand synthetic images featuring diverse hand shapes and backgrounds [22]. Thus, the model demonstrates the potential to produce high-quality classifications in various settings. For classification, different ML techniques exist, e.g., Artificial Neural Networks (ANN), Support Vector Machines (SVM), Decision Trees (DT), and Naïve Bayes Classifiers (NBC) [38]. ANN and SVM are among the most commonly used classifiers in the literature [39].

Hand gesture recognition is an emerging interaction modality for human-CPS interactions. By exploring different approaches, including wearable sensors, RGB-D sensors, and CNNs, researchers have made progress in accurate and efficient hand gesture recognition. Transfer learning offers viable strategies for improving the performance of hand gesture recognition models by using pre-trained models like Mediapipe Hands to further enhance the accuracy and speed of recognition. These advancements contribute to the growing body of knowledge on hand gesture recognition as a practical and effective modality for human-CPS interaction.

**Voice commands as an interaction modality**

Voice command systems have been introduced for many applications in both homes as well as workplaces. These systems are used for providing cognitive support (e.g., simple calculations), controlling devices, and hands-free access to information [15]. Some commercial examples of voice command systems are voice assistants like Google Home, Amazon Alexa, Apple Siri, etc. [15]. It is estimated that 27% of the online population uses voice search and 17% own a voice assistant [23]. In the built environment, these voice assistants are used in smart homes or smart office applications for controlling devices and home appliances [23,40]. In construction, voice command systems have been studied to provide hands-free information about construction assembly to workers for faster construction time due to reduced time in searching for information in paper documents (drawings) [15]. However, these commercial systems have limited vocabulary and also require additional dedicated hardware. There is also a lack of research on interoperability between these voice assistants and construction technologies like BIM [15].

Elghaish et al. [41] integrated voice assistants with BIM for data management. Voice-based interactions with BIM allow even novice users to perform advanced tasks like parametric design and can even help disabled users perform certain tasks that might not be possible for them with a mouse and keyboard [41]. Similarly, Linares-Garcia et al. [15] used voice assistants to extract assembly information from BIM and provide construction workers with hands-free access to information. However, both of the authors used Alexa and Google Home devices respectively for recognizing voice commands. These voice assistant devices are general-purpose hardware that must be initiated with specific wake words [42]. They are also not integrated with the CPS. Therefore, in this study, a methodology is developed to directly recognize speech to provide input for a CPS for better integration of the interaction modality with the CPS without the need for additional hardware.

**Research Gaps**

Even though hand gestures have been studied as an interaction modality for human-CPS interaction in past studies [43], past research has used either wearable sensors to recognize hand gestures or deep learning methods like CNN. Each of these methods has significant practical constraints. For example, wearable sensors require additional expensive hardware which can be uncomfortable to wear [21]. On the other hand, deep learning methods require large datasets to train [21]. The transfer learning approach has not been used for hand gesture recognition, specifically in the context of human-CPS interaction in construction.

Additionally, voice-based systems are gaining popularity in homes as well as workplaces. However, there is a scarcity of research on voice-based systems in construction. Few research that has been conducted in the past utilized voice assistant devices like Google Home and Amazon Alexa [15,41]. These voice assistants require additional dedicated hardware and are also not well-integrated with the CPS.

Therefore, this study develops two methodologies for human-CPS interactions with hand gesture and voice command recognition. These methodologies will fill the gap in research on these interaction modalities for human-CPS interactions, especially in the context of construction.

## METHODOLOGY

### Overall Research Methodology

The overall methodology of this study is shown in Figure 1. First, the state-of-the-art of hand gestures and voice command interactions were reviewed from the literature. Based on the findings from the literature review, gaps were identified for utilizing hand gestures and voice commands in human-CPS interaction. An ML pipeline was developed for hand gesture recognition using a transfer learning approach. Then, a voice recognition model was developed that records and extracts voice inputs by the user using a microphone and matches them with a list of commands to recognize voice commands by calculating similarities between the recorded input and each command in the command list. The details of the developed methodologies for hand gesture and voice command recognition are presented in Sections 0 and 0 respectively.

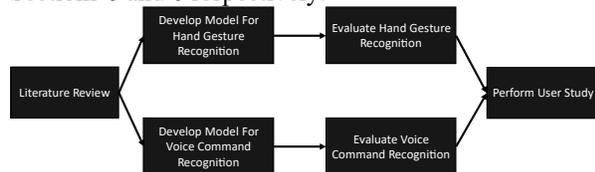

Figure 1 Study Methodology

Both models were evaluated separately by generating a labeled validation dataset with hand images and voice snippets. The video dataset generated for hand gesture recognition was split into training and test datasets in an 80/20 ratio, i.e., only 80% of the data was used for training the model and the rest 20% was kept aside for testing and evaluating the model. The hand gesture recognition model was evaluated on the following criteria (a to f):

a) Training time – Total time taken for training the model.
b) Accuracy – Ratio of inputs incorrectly classified by the model out of all inputs tested. Mathematically written as
$Accuracy = True\ Positives/Number\ of\ inputs$
c) Precision - Ratio of inputs correctly classified as one of the gestures to the inputs classified as the same gesture (including wrong predictions) averaged over all the gestures. Mathematically written as:
$Precision = avg_{classes}[True\ Positives/(True\ Positives + False\ Positives)]$
d) Recall – Ratio of inputs correctly classified as one of the gestures to the number of inputs for that gesture averaged over all the gestures. Mathematically written as:
$Recall = avg_{classes}[True\ Positives/(True\ Positives + False\ Negatives)]$
e) F1 score – Harmonic mean of precision and recall. Mathematically written as:
$F1 = Precision * Recall/(Precision + Recall)$
f) Prediction time – Average time taken by the model to recognize each input.

Voice command recognition as explained in Section 5 did not have a training involved. Therefore, it was evaluated on the criteria b to f above. After evaluating and fine-tuning the two interaction modalities, a user study was conducted to evaluate the potential of the two input modalities for human-CPS interactions for construction progress monitoring. Details of the user study is provided in Section 0.

### User Study Methodology

To evaluate the effectiveness of the proposed hand gesture and voice command recognition systems for construction progress monitoring and test the generalizability of the two methods for a wider user population, a user study was conducted with 32 participants with a background in construction. The participants were asked to use a desktop-based progress monitoring system to monitor the progress of a construction site using only hand gestures and voice commands as the interaction modalities. The experimental setup is shown in Figure 2. A pilot test was also conducted with two participants, whose data is not included in the analysis. The purpose of the pilot test was to refine the instructions and test procedures for the included participants. The eligibility criteria for the participants were set as self-reported familiarity with construction schedules and construction drawings, an age of at least 18 years to provide consent, and a normal or corrected-to-normal vision. The whole experiment protocol was reviewed and approved by Virginia Tech Institutional Review Board (IRB) under protocol #22-978.

During the study, the participants were asked to complete a construction progress monitoring task by identifying the construction activities that were completed on a given schedule. The term *task* refers to the progress monitoring task that the users were instructed to perform with the CPS, which involved identifying the completed *activities* on the schedule. The schedule provided to the users for the progress monitoring task is presented in Appendix A. They marked the activities as complete or incomplete on a printed copy of the schedule based on their observation of the work through the visualized site images. As shown in Figure 2, the user sat in front of a monitor which showed the interface of the CPS. Their hand gestures were recorded through a webcam installed on top of the monitor which was used to send inputs to the CPS. During the voice command interaction, users provided voice inputs to the CPS to perform the progress monitoring task. The voice commands were provided through a JBL Bluetooth speaker and microphone kept in front of the user.

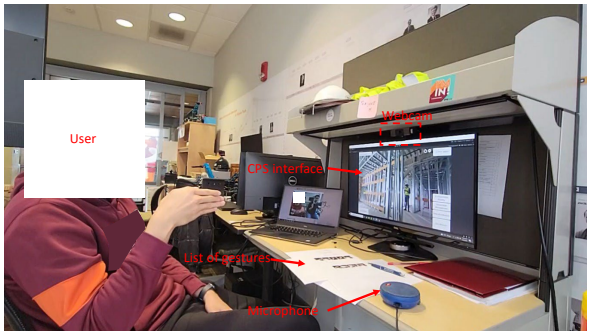

Figure 2 A user interacting with a CPS for progress monitoring using hand gestures

The progress monitoring CPS used for the user study consisted of a virtual environment created from a BIM. The CPS integrates BIM, reality capture, and robot control to autonomously capture 360° images from a construction site and visualize them in the virtual environment geolocated at the location of capturing images. Figure 3 shows the relationship between different components of the CPS. The CPS is an autonomous system which requires user input for locations where the user wants to monitor progress. Based on the information provided, the system navigates the robot through the project site. The robot carries a 360° camera that captures reality from the site and sends back to the interface for visualization. Readers are referred to the author's previous paper [9] for more details regarding the implementation of the CPS.

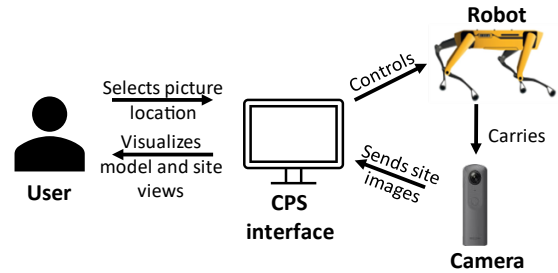

Figure 3 Relationships between the components of the CPS

The interface of the CPS is shown in Figure 4. In the virtual environment, the users can navigate the BIM model of the whole project area, look around the model, enter reality points that represent the 360° pictures captured by the robot to see site visuals, and also see the floor plan and schedule of the project. The model and site images used were captured by the researchers from a live construction project using an autonomous quadruped robot called Spot and were pre-loaded on the CPS interface. Users only interacted with the CPS interface to view the model, site images, floor plan, and schedule of the project to determine what construction activities from the schedule were completed. The project was the new construction of an apartment building on the Virginia Tech campus in Blacksburg, USA with a gross area of about 2949 square feet. In total, thirteen 360° images were used from the construction site. At the time of capturing images, structural framing and curtain walls were already constructed and partition walls were being constructed.

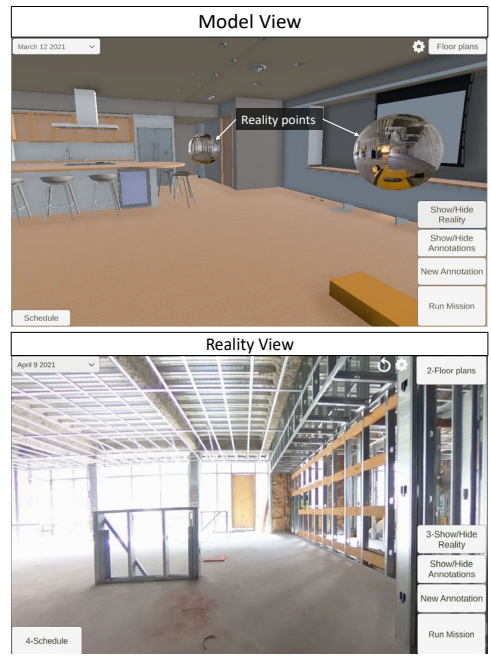

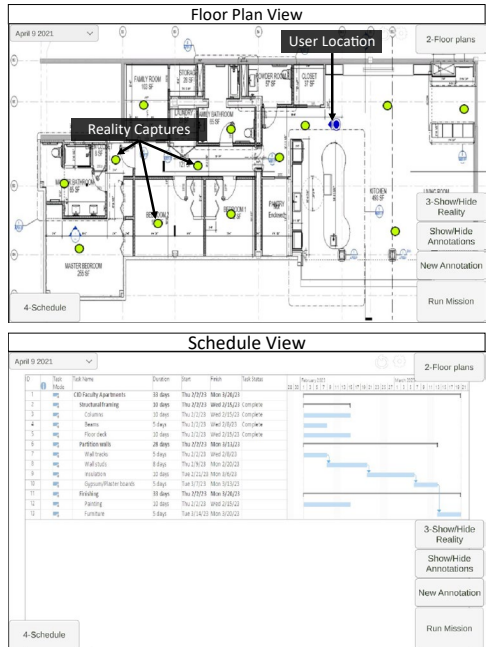

Figure 4 CPS interface used for progress monitoring in the user study

The participants were instructed to navigate through the model and images by using hand gestures and voice commands in two different tasks. Possible commands that the CPS can understand are shown in Figure 5 below. The allowed input commands for the CPS are explained in Sections 0 and 0. They could navigate the model by moving and rotating the view. They could also show/hide the floor plan and schedule of the project and show/hide the reality points so that they can navigate the BIM model easily without getting distracted by the reality points. The participants were given a brief training session on how to use the hand gesture recognition system. A cheatsheet was provided to the users in printed form to avoid the need for memorizing the gestures and commands. After training, users were allowed to practice until they felt confident in using the system with interaction modality. They were also allowed to ask any questions they had during practice. Total training and practice time ranged from 5 to 10 minutes approximately for each user participant. No time limit was placed on the training and practice. However, for performing the progress monitoring task, users were allowed only ten minutes, though most users completed the task in less time than that. The data were analyzed using descriptive statistics. The number of activities marked on the schedule by the users and the number of activities that were marked correctly were used as a measure of completeness and correctness respectively and were counted manually from the schedule by the researcher after the experiments.

## VISION-BASED HAND-GESTURE RECOGNITION

This section describes the technical details of the hand gesture recognition system developed in this study. The prototype system used Mediapipe Hands as a feature generator and a Support Vector Machine (SVM) with a linear kernel and an ANN with 30 hidden neurons as classifiers. The SVM was trained using the scikit-learn package in Python [44], whereas the ANN was trained with the TensorFlow library [45]. The video dataset was generated by recording videos of the researcher making different hand gestures using a webcam. A total of 15 gestures were recorded that reflect the different functionalities of the remote progress monitoring system used for the user study as shown in Figure 5. The look gestures in Figure 5 controlled the rotation of the view in the direction users pointed with their index fingers. The move gestures were used to move/walk around the model and move the view in the direction the users pointed with all their fingers. In addition, show/hide gestures could be used to show/hide the floor plan, reality, and schedule. The floor plan, reality, and schedule were assigned the numbers two, three, and four respectively. This means that making a gesture for number 2 with the palm side of the hand as shown in Figure 5 would show the floor plan and making the gesture for number 2 with the back of the hand labeled as reverse_two would hide the floor plan. Similarly, the reality points (the 360° images) and the schedule could be shown or hidden with the gestures for number 3 and number 4 respectively. Videos for each gesture were recorded for 40 seconds at 25 frames per second, thereby, producing 1000 frames per gesture. One of the emphases of the study was also to make the ML pipeline scalable by keeping the training process simple and fast, i.e., more gestures can be added as needed by re-training the classifier. As is explained in the next subsection, the classifier took only 1.36 seconds to train. Therefore, the gestures can be added by re-training the model as and when required.

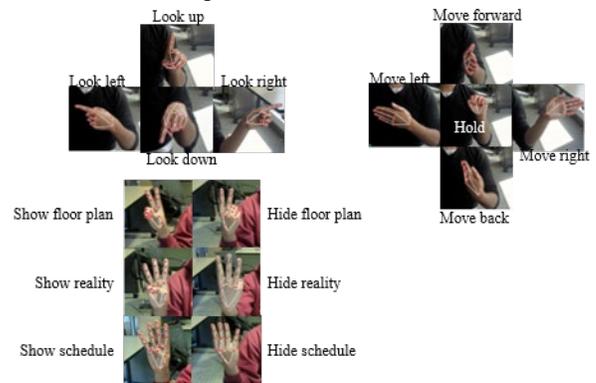

Figure 5 Hand gestures used for human-CPS interaction

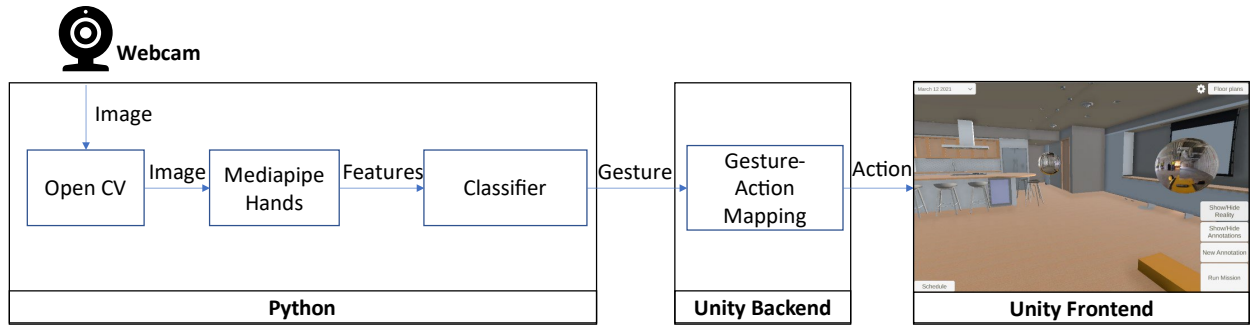

Figure 6 Proposed ML pipeline for hand-gesture recognition as an input for the studied CPS

The recorded videos are then processed through Mediapipe to estimate landmark positions which are stored in an in-memory 2D dataframe in Pandas, a data processing library in Python. Each row in the 2D dataframe consists of 63 values representing the cartesian coordinates (x, y, and z) of 21 landmarks on the right hand of a human. The labels are stored in another 1D dataframe in Pandas. Before training, the 1D dataframe is encoded using the One-Hot encoding method, in which each unique label is represented by a column with binary values as shown in Figure 7, essentially converting the labels from textual representation into a trainable numerical representation. The compiled numerical representation of the features is then passed through an ANN and an SVM classifier. The ANN model comprised 30 neurons in the hidden layer and 63 neurons in the input layer. The output layer consisted of a number of neurons that equals the number of gestures to be trained. The SVM classifier has a much smaller number of customizable hyper-parameters. A simple SVM model with a linear kernel and the regularization parameter C as 1 was used for training which was later found to be adequate.

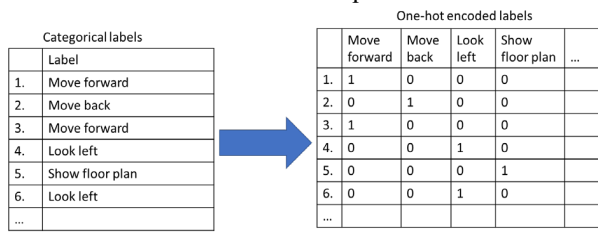

Figure 7 One-hot encoding

During runtime, the prototype system captured video frames of the user's hand using a webcam and processed the video stream frame-by-frame through Mediapipe to extract the hand landmarks coordinates. These landmarks coordinates are then passed as features to the trained SVM and ANN classifiers, which predict the hand gesture label based on the learned decision boundaries. The predicted gesture is then sent to the construction progress monitoring system as an input command.

## SPEECH RECOGNITION FOR VOICE COMMANDS INTERACTION

This section provides the technical details of the proposed method of speech recognition for voice command interaction between humans and CPS. As shown in Figure 8, the speech input is received from the user through a microphone. The audio input is recorded in three-second intervals. The three-second snippet of audio is then transcribed using the speech-to-text service in Google Cloud Platform (GCP). The GCP returns the transcribed text for the provided audio snippet. The received text is then passed through a Speech Recognition Module (SRM) developed in Python. The module calculates the similarity of the transcribed phrase with the phrases in a command list, which represents the expected commands. Two similarity scores are calculated between the transcribed phrase and the expected commands – Cosine similarity and Jaro-Winkler similarity. Cosine similarity is the measure of the angle between the word vector representations of the two text phrases. A word vector is a numerical representation of a word in a multi-dimensional feature space. Two words will have similar word vectors when they are close in meaning and are often used in the same context, for example, move and walk. The spaCy package for Python [46] provides pre-trained word vectors for more than 600,000 words, which was used in this study for speech recognition. On the other hand, the Jaro-Winkler similarity is the measure of edit distance between the two text phrases, i.e., how many edits need to be made to convert one phrase to another [47]. Jaro-Winkler similarity measure is very efficient and flexible in finding similarities between two text phrases [47].

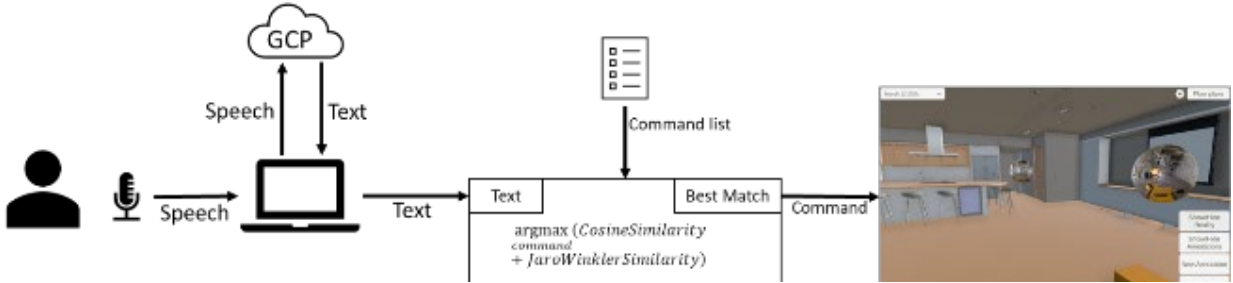

Figure 8 Overview of the proposed method of speech recognition for voice command interaction between humans and CPS.

The sum of the Cosine similarity and Jaro-Winkler similarity is considered the total similarity of two phrases which is then used to find the best match from the command list with the input command. Both the Cosine similarity and Jaro-Winkler similarity range from 0 to 1. The command from the command list with maximum total similarity is considered as the recognized command if the total similarity of the command is more than 1. If none of the commands has a total similarity of more than 1, the speech input is ignored.

## RESULTS

### Evaluation of Hand Gesture Recognition Methodology

The observed values for the evaluation metrics explained in Section 0 are reported in Table 1. As can be seen from the table, SVM was found to be better than ANN on all the metrics. Therefore, SVM was chosen as a classifier in the proposed ML pipeline. Figure 9 shows the confusion matrix for the SVM classifier, which compares the true labels with the predicted labels by the classifier for each of the classes. As can be seen from the confusion matrix, the classifier was able to classify the gestures correctly in almost all cases. The evaluation was performed with 20% of the dataset preserved for testing and not used for training. A lower accuracy with unseen data than the accuracy with training data would indicate overfitting [38]. However, the accuracy of the model with the unseen test data was found to be high (99.69%). This suggested no over-fitting and therefore further tuning of the model was not performed which would be required if overfitting were observed.

Table 1 Evaluation metrics for the hand gesture classifiers

| Metric | SVM | ANN |
|---|---|---|
| Training time | 1.36s | 4.54s |
| Accuracy | 99.69% | 95.10% |
| Precision | 99.62% | 94.68% |
| Recall | 99.66% | 94.22% |
| F1 score | 99.64% | 94.37% |
| Prediction time | 36.05 ms/prediction | 181.84 ms/prediction |

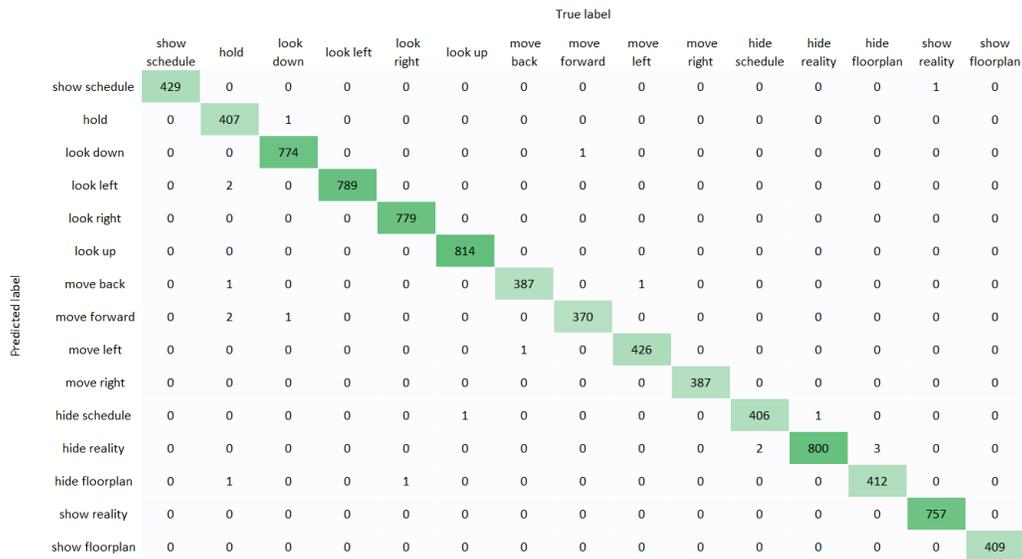

Figure 9 Confusion matrix for hand gesture classification system

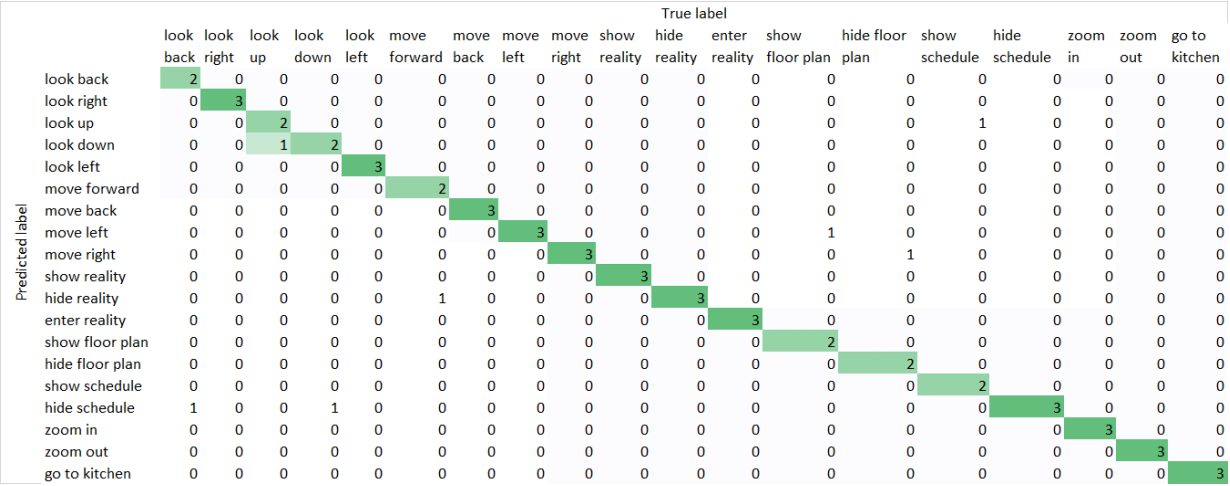

Figure 10 Confusion matrix for voice command recognition system

**Evaluation of Voice Command Recognition Methodology**

The proposed voice command recognition system was evaluated using the same metrics as the hand gesture recognition system, i.e., accuracy, precision, recall, f1 score, and prediction time. Since there is no model training involved in the proposed methodology, training time was not used as an evaluation criterion. A validation dataset was generated by recording voice snippets for each command (look back, look right, look up, look down, look left, move forward, move back, move left, move right, show reality, hide reality, enter reality, show floor plan, hide floor plan, show schedule, hide schedule, zoom in, zoom out, go to kitchen) by the author. The commands in semantically similar utterances, for example, for "move forward" utterances like "go forward" and "walk forward" were also included in the validation dataset. Each command was repeated three times in the test dataset including different utterances for the same command, thereby generating 57 (19 * 3) voice snippets. The results of the evaluation of the proposed voice command recognition system are provided in Table 2. By comparing Table 1 and Table 2, it can be seen that the proposed hand gesture recognition system performed better than the proposed voice command recognition system in all the metrics. This can be attributed to the fact that human voices have much more diversity than human hands. Also, an image provides more information than a voice which helps the prediction. Figure 10 shows the confusion matrix generated from the evaluation of the voice command system.

Table 2 Evaluation metrics for the voice command recognition system

| Metric | Value |
| --- | --- |
| Accuracy | 87.72% |
| Precision | 90.44% |
| Recall | 87.72% |
| F1 score | 87.66% |
| Prediction time | 578.91 ms/prediction |

**User study results**

The demographics of the user participants are shown in Figure 11. The average age of the participants was 27.6 years. Out of 32 participants, 22 (69%) were male and 10 (31%) were female.

The users were allowed a maximum of ten minutes to complete the progress monitoring tasks. The time of completion of the task was also noted as well as the completeness and correctness of the task of identifying the completed and incomplete activities on the schedule.

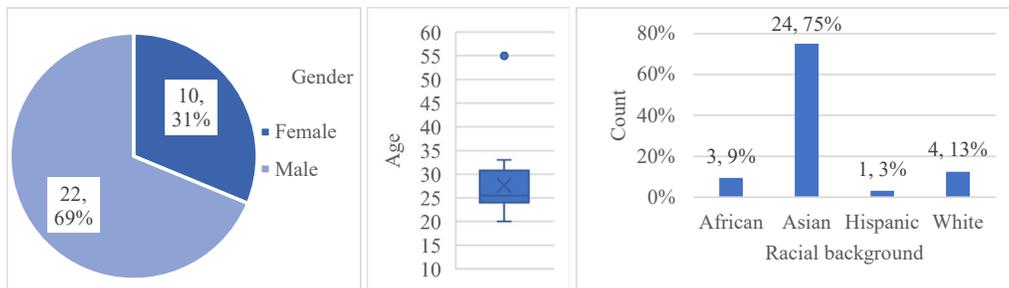

Figure 11 Demographics of the user participants

As can be seen from Figure 12 (left), most participants (30 out of 32 or 93.75%) were able to complete all the tasks in the given ten minutes with the hand gesture interaction. 29 out of 32 users (90.62%) were able to complete the progress monitoring tasks within the given ten minutes with the voice command interaction. The participants completed the tasks with an average accuracy of 66.1% and 88.0% for hand gesture and voice command interaction. Figure 12 shows the histogram of the number of tasks completed and correctly completed by the participants with each of the interaction modalities.

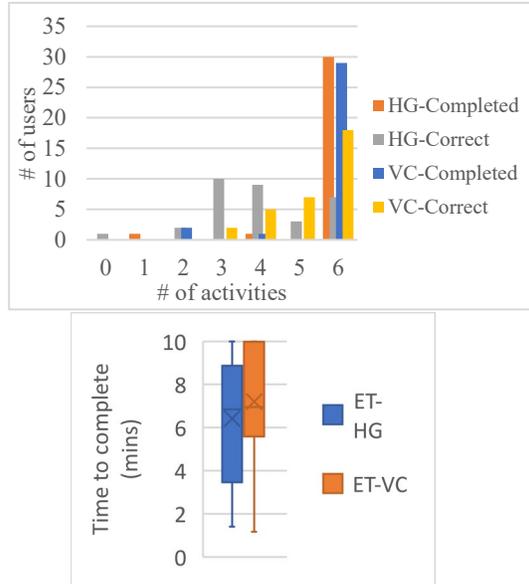

Figure 12 Task completeness and correctness (left) and task completion time (right) with hand gesture and voice command control

## DISCUSSION

The results of the user study demonstrated the potential of using hand gestures and voice commands as alternative human-CPS interaction modalities in a use case of remote construction progress monitoring. 93.75% and 90.62% users were able to complete the given progress monitoring task of updating a construction schedule in a limited timeframe using hand gestures and voice command recognition using the proposed methodologies respectively. This shows the potential of using these interaction modalities for human-CPS interaction.

From the results of the validation tests performed with test data, it was found that even though ANN is a more complex model with the ability to classify input in a complex non-linear feature space, it also has more training parameters and therefore is slower than SVM for the hand gesture classification task. With the given feature space, SVM even though a simpler model than ANN was found to be more accurate with faster training and prediction. The prototype design for the hand gesture recognition system, which used Mediapipe Hands as a feature generator and Support Vector Classifier with a linear kernel for classification, achieved a high accuracy rate of 99.69% on the test dataset. The use of the linear kernel can be beneficial for real-time applications, as it has faster training and prediction time compared to non-linear kernels while providing good accuracy. The model was able to perform at a rate of 36.05 ms/prediction, which was found to be adequate for real-time interaction with the remote monitoring system in the user studies.

The voice command recognition system based on converting speech to text and calculating similarities between the converted text and a list of commands was found to be a fast and accurate method of using voice commands as a human-CPS interaction modality. Past research with voice command systems in construction [15,41] used dedicated voice assistant devices like Alexa and Google Home. Using the proposed method, the interaction modality can be integrated within the CPS and the need for a dedicated voice assistant device can be eliminated. However, voice command recognition with the proposed methodology was found to be slower and less accurate than the proposed hand gesture recognition system. One of the advantages of using voice command systems is that it provides a hands-free method of interacting with a computer system thereby allowing the users to multi-task. However, it should be noted that not all tasks might be suitable to perform with voice command systems. One of the limitations of using voice commands found in this study is that it takes considerable time to extract voice commands from speech inputs. As was found in the evaluation, the process of recognizing the voice command using the proposed methodology took about 579 ms which is slower than other interaction modalities like hand gestures which took about 36 ms to recognize input. Therefore, more research is needed to identify tasks that can benefit from voice commands as an input modality.

The participants completed the tasks with an average correctness of 66.1% and 88.0% for hand gesture and voice command interaction, respectively. This finding suggests that participants achieved higher accuracy in the given progress monitoring task when utilizing voice commands compared to hand gestures. Even though the validation tests indicated higher accuracy for hand gesture recognition using the proposed methodologies. This divergence between participant accuracy and predictive performance could be attributed to factors such as the participants' familiarity and comfort with using voice commands compared to hand gestures. Participants may have had more experience with voice command interactions (e.g., with voice assistants like Siri and Alexa) in their daily lives, leading to greater accuracy in executing the given tasks. These findings underscore the importance of considering user preferences and experience when designing and implementing interaction modalities in CPS.

Nonetheless, despite the differing accuracy rates, both hand gesture and voice command interaction modalities demonstrated potential for human-CPS interaction in the context of construction progress monitoring. The high accuracy achieved by the proposed methodology for hand gesture recognition, and the convenience and hands-free nature of voice commands, also indicated the potential of these modalities to be used in conjunction with each other for more efficient human-CPS interaction.

## CONCLUSION

In this study, the potential of hand gesture and voice command recognition was explored as alternative human-CPS interaction modalities in a use case of construction progress monitoring. A prototype hand gesture recognition system was designed using a feature representation transfer learning approach. Both SVM and ANN were tested as classifiers. It was found that SVM with a linear kernel provided high accuracy as well as training and prediction rates. The trained model was evaluated with a validation dataset. The accuracy achieved with the proposed methodology for hand gesture recognition on the validation dataset was 99.69%. The model training was completed in 1.36 s with 15 gestures. The model also performed at a rate of 36.05 ms which was found to be fast enough for completing a progress monitoring task in the user studies. The proposed hand gesture recognition model uses a landmark-based approach that analyzes the relative positions of 21 different landmarks on a human hand. As compared to the pixel-level approach used by past researchers, the landmark-based approach significantly reduces the feature space, thereby, reducing the training time and improving prediction speed, which is important for real-time human-CPS interaction.

Similarly, a methodology for recognizing voice commands for human-CPS interaction was also developed in this study. The proposed methodology works by first recording and converting a speech input by the user into text. Then, comparing the converted text with a list of functionalities the CPS can perform. The proposed methodology was validated with a validation dataset generated by the author by recording voice snippets for each command the CPS can understand. The proposed methodology was found to have a reasonable accuracy of 87.72% and a prediction rate of 578.91 ms which was slower than hand gesture recognition but was found to be fast enough to perform a progress monitoring task.

The strength of both methods lies in the elimination of the need for large training data for hand gestures and voice command recognition. To evaluate the generalizability of the two methodologies trained with a small training dataset, a user study was conducted with 32 participants with a background in construction. In the user study, the user participants used a desktop-based CPS interface to monitor construction progress using hand gestures and voice commands as input modalities. The results of the study showed that both hand gesture and voice command recognition systems achieved high accuracy and real-time operation which allowed the users to complete the given progress monitoring task. 93.75% and 90.62% of users were able to complete the task in the given time with hand gestures and voice commands respectively.

Limitations of the study include a controlled environment for the evaluation of the system. The accuracy and effectiveness of the system should be tested in a crowded environment as well. The user study was also executed in a short timeframe. The learnability of hand gestures and voice commands as input modalities and the effect of more training on the effectiveness of the proposed methodologies can be studied in longer studies.

Future work can further refine and develop the proposed interaction modalities, exploring their integration with other CPS technologies, such as drones and robots in a construction environment. Further research can also investigate the use of these interaction modalities in other use cases, such as design, facility management, and smart building operations. For hand gesture recognition, two of the most commonly used classifiers ANN and SVM were used and compared. Future research can also explore other classification methods like Decision Tree, Random Forest, and K-means. Finally, during the experiments, users pointed out the benefits and challenges of each of the interaction modalities. Therefore, a multimodal human-CPS interaction that combines both of these interaction modalities may be explored.

In conclusion, the results of this study suggest that hand gesture and voice command recognition can be potential alternative modalities for human-CPS interaction in construction, particularly in construction progress monitoring. The hand gesture recognition system, using Mediapipe Hands as a feature generator and Support Vector Machine with a linear kernel, achieved high accuracy and user participants were able to use it in real time. The voice recognition system by combining the Cosine similarity and Jaro-Winkler similarity to recognize voice commands was shown as a potential method of integrating voice recognition within a CPS. Further research and development of these technologies can potentially improve the efficiency, effectiveness, and user satisfaction of using CPS in construction.

## ACKNOWLEDGMENTS

This study was partially funded by the Industrialized Construction Initiative at Virginia Tech College of Engineering. Additionally, the authors acknowledge the work of undergraduate student Kalen Rita for supporting this research.

## REFERENCES

[1] A. Sawhney, M. Riley, J. Irizarry, eds., Construction 4.0: An Innovation Platform for the Built Environment, 2020. https://www.routledge.com/Construction-40-An-Innovation-Platform-for-the-Built-Environment/Sawhney-Riley-Irizarry/p/book/9780367027308 (accessed May 15, 2023).

[2] J.J. Lin, M. Golparvar-Fard, Construction Progress Monitoring Using Cyber-Physical Systems, in: C. Anumba, N. Roofigari-Esfahan (Eds.), Cyber-Physical Systems in the Built Environment, 2020: pp. 63–87. https://doi.org/10.1007/978-3-030-41560-0.

[3] S. Eskandar, J. Wang, S.N. Razavi, Human-in-the-Loop Cyber-Physical Systems for Construction Safety, in: Cyber-Physical Systems in the Built Environment, 2020: pp. 161–173.

[4] X. Yuan, C.J. Anumba, Cyber-Physical Systems for Temporary Structures Monitoring, in: C.J. Anumba, N. Roofigari-Esfahan (Eds.), Cyber-Physical Systems in the Built Environment, Springer International Publishing, Cham, 2020: pp. 107–138. https://doi.org/10.1007/978-3-030-41560-0_7.

[5] Y. Fang, Y.K. Cho, C. Kan, C.J. Anumba, Cyber-Physical Systems (CPS) in Intelligent Crane Operations, Cyber-Physical Systems in the Built Environment. (2020) 175–192. https://doi.org/10.1007/978-3-030-41560-0_10.

[6] S. Jeong, R. Hou, J.P. Lynch, K.H. Law, Structural-Infrastructure Health Monitoring, in: C.J. Anumba, N. Roofigari-Esfahan (Eds.), Cyber-Physical Systems in the Built Environment, Springer International Publishing, Cham, 2020: pp. 215–235. https://doi.org/10.1007/978-3-030-41560-0_12.

[7] J. Wu, D. Fang, Role of CPS in Smart Cities, in: C.J. Anumba, N. Roofigari-Esfahan (Eds.), Cyber-Physical Systems in the Built Environment, Springer International Publishing, Cham, 2020: pp. 255–272. https://doi.org/10.1007/978-3-030-41560-0_14.

[8] S. Terreno, A. Akanmu, C.J. Anumba, J. Olayiwola, Cyber-Physical Social Systems for Facility Management, in: C.J. Anumba, N. Roofigari-Esfahan (Eds.), Cyber-Physical Systems in the Built Environment, Springer International Publishing, Cham, 2020: pp. 297–308. https://doi.org/10.1007/978-3-030-41560-0_16.

[9] S. Halder, K. Afsari, J. Serdakowski, S. DeVito, A Methodology for BIM-enabled Automated Reality Capture in Construction Inspection with Quadruped Robots, in: Proceedings of the International Symposium on Automation and Robotics in Construction, 2021: pp. 17–24. https://www.scopus.com/inward/record.uri?eid=2-s2.0-85127582547&partnerID=40&md5=7136475af2eb00129ce03f510c34b960.

[10] M. Kopsida, I. Brilakis, P. Vela, A Review of Automated Construction Progress and Inspection Methods, in: Proceedings of the 32nd CIB W78 Conference on Construction IT, 2015: pp. 421–431.

[11] D. Gorecky, M. Schmitt, M. Loskyll, D. Zühlke, Human-machine-interaction in the industry 4.0 era, in: 2014 12th IEEE International Conference on Industrial Informatics (INDIN), Ieee, 2014: pp. 289–294.

[12] S. Ahmed, V. Popov, A. Topalov, N. Shakev, Hand Gesture based Concept of Human - Mobile Robot Interaction with Leap Motion Sensor, IFAC-PapersOnLine. 52 (2019) 321–326. https://doi.org/10.1016/j.ifacol.2019.12.543.

[13] C. Krupitzer, S. Müller, V. Lesch, M. Züfle, J. Edinger, A. Lemken, D. Schäfer, S. Kounev, C. Becker, A Survey on Human Machine Interaction in Industry 4.0, (2020). https://doi.org/10.48550/arXiv.2002.01025.

[14] X. Wang, Z. Zhu, Vision-based hand signal recognition in construction: A feasibility study, Automation in Construction. 125 (2021) 103625. https://doi.org/10.1016/j.autcon.2021.103625.

[15] D.A. Linares-Garcia, N. Roofigari-Esfahan, K. Pratt, M. Jeon, Voice-Based Intelligent Virtual Agents (VIVA) to Support Construction Worker Productivity, Automation in Construction. 143 (2022) 104554. https://doi.org/10.1016/j.autcon.2022.104554.

[16] S. Khan, B. Tunçer, Gesture and speech elicitation for 3D CAD modeling in conceptual design, Automation in Construction. 106 (2019) 102847. https://doi.org/10.1016/j.autcon.2019.102847.

[17] C. Bassano, M. Chessa, F. Solari, A Study on the Role of Feedback and Interface Modalities for Natural Interaction in Virtual Reality Environments., in: VISIGRAPP (2: HUCAPP), 2020: pp. 154–161.

[18] Z. Xia, Q. Lei, Y. Yang, H. Zhang, Y. He, W. Wang, M. Huang, Vision-based hand gesture recognition for human-robot collaboration: A survey, in: 2019 5th International Conference on

Control, Automation and Robotics (ICCAR), IEEE, 2019: pp. 198–205.
[19] A.S. Al-Shamayleh, R. Ahmad, M.A.M. Abushariah, K.A. Alam, N. Jomhari, A systematic literature review on vision based gesture recognition techniques, Multimedia Tools and Applications. 77 (2018) 28121–28184.
[20] F. Liu, B. Du, Q. Wang, Y. Wang, W. Zeng, Hand gesture recognition using kinect via deterministic learning, in: 2017 29th Chinese Control and Decision Conference (CCDC), IEEE, 2017: pp. 2127–2132.
[21] Z. Kolar, H. Chen, X. Luo, Transfer learning and deep convolutional neural networks for safety guardrail detection in 2D images, Automation in Construction. 89 (2018) 58–70.
[22] F. Zhang, V. Bazarevsky, A. Vakunov, A. Tkachenka, G. Sung, C.-L. Chang, M. Grundmann, MediaPipe Hands: On-device Real-time Hand Tracking, ArXiv:2006.10214. (2020). http://arxiv.org/abs/2006.10214.
[23] C. Jimenez, E. Saavedra, G. del Campo, A. Santamaria, Alexa-Based Voice Assistant for Smart Home Applications, IEEE Potentials. 40 (2021) 31–38. https://doi.org/10.1109/MPOT.2020.3002526.
[24] A.A. Akanmu, C.J. Anumba, O.O. Ogunseiju, Towards next generation cyber-physical systems and digital twins for construction, Journal of Information Technology in Construction. 26 (2021) 505–525. https://doi.org/10.36680/j.itcon.2021.027.
[25] B. Dafflon, N. Moalla, Y. Ouzrout, The challenges, approaches, and used techniques of CPS for manufacturing in Industry 4.0: a literature review, The International Journal of Advanced Manufacturing Technology. 113 (2021) 2395–2412. https://doi.org/10.1007/s00170-020-06572-4.
[26] Y. Srewil, A. Ismail, R.J. Scherer, A Method to Integrate Virtual-Physical Construction Environment in Framework of CPS Approach, in: Smart SysTech 2016; European Conference on Smart Objects, Systems and Technologies, 2016: pp. 1–8.
[27] M. Gil, M. Albert, J. Fons, V. Pelechano, Engineering human-in-the-loop interactions in cyber-physical systems, Information and Software Technology. 126 (2020) 106349. https://doi.org/10.1016/j.infsof.2020.106349.
[28] F. R. Correa, Cyber-physical systems for construction industry, in: 2018 IEEE Industrial Cyber-Physical Systems (ICPS), 2018: pp. 392–397. https://doi.org/10.1109/ICPHYS.2018.8387690.
[29] D.A. Linares Garcia, N. Roofigari-Esfahan, Technology requirements for CPS implementation in construction, in: Cyber-Physical Systems in the Built Environment, Springer, 2020: pp. 15–30.
[30] B.M. Tehrani, J. Wang, C. Wang, Review of Human-in-the-Loop Cyber-Physical Systems (HiLCPS): The Current Status from Human Perspective, in: Computing in Civil Engineering, 2019: pp. 470–478. https://doi.org/10.1061/9780784482438.060.
[31] D. Nunes, J.S. Silva, F. Boavida, A Practical Introduction to Human-in-the-loop Cyber-physical Systems, John Wiley & Sons, 2018.
[32] F. Tao, Q. Qi, New IT Driven Service-Oriented Smart Manufacturing: Framework and Characteristics, IEEE Transactions on Systems, Man, and Cybernetics: Systems. 49 (2019) 81–91. https://doi.org/10.1109/TSMC.2017.2723764.
[33] D. Mukherjee, K. Gupta, L.H. Chang, H. Najjaran, A Survey of Robot Learning Strategies for Human-Robot Collaboration in Industrial Settings, Robotics and Computer-Integrated Manufacturing. 73 (2022) 102231. https://doi.org/10.1016/J.RCIM.2021.102231.
[34] D. Droeschel, J. Stückler, S. Behnke, Learning to interpret pointing gestures with a time-of-flight camera, in: 2011 6th ACM/IEEE International Conference on Human-Robot Interaction (HRI), 2011: pp. 481–488. https://doi.org/10.1145/1957656.1957822.
[35] C. Nuzzi, S. Pasinetti, R. Pagani, S. Ghidini, M. Beschi, G. Coffetti, G. Sansoni, MEGURU: a gesture-based robot program builder for Meta-Collaborative workstations, Robotics and Computer-Integrated Manufacturing. 68 (2021) 102085.
[36] F. Kucuksubasi, A.G. Sorguc, Transfer Learning-Based Crack Detection by Autonomous UAVs, in: J. Teizer (Ed.), Proceedings of the 35th International Symposium on Automation and Robotics in Construction (ISARC), IAARC, Taipei, Taiwan, 2018: pp. 593–600.
[37] S.J. Pan, Q. Yang, A Survey on Transfer Learning, IEEE Transactions on Knowledge and Data Engineering. 22 (2010) 1345–1359. https://doi.org/10.1109/TKDE.2009.191.
[38] A. Géron, Hands-On Machine Learning with Scikit-Learn, Keras, and TensorFlow, 2019.
[39] I. Peško, V. Mučenski, M. Šešlija, N. Radović, A. Vujkov, D. Bibić, M. Krklješ, Estimation of Costs and Durations of Construction of Urban Roads Using ANN and SVM, Complexity. 2017 (2017) e2450370. https://doi.org/10.1155/2017/2450370.
[40] Z.Y. Chan, P. Shum, Smart Office: A Voice-controlled Workplace for Everyone, in: Proceedings of the 2nd International Symposium

on Computer Science and Intelligent Control, Association for Computing Machinery, New York, NY, USA, 2018: pp. 1–5. https://doi.org/10.1145/3284557.3284712.

[41] F. Elghaish, J.K. Chauhan, S. Matarneh, F. Pour Rahimian, M.R. Hosseini, Artificial intelligence-based voice assistant for BIM data management, Automation in Construction. 140 (2022) 104320. https://doi.org/10.1016/j.autcon.2022.104320.

[42] I. Motawa, Spoken dialogue BIM systems – an application of big data in construction, F. 35 (2017) 787–800. https://doi.org/10.1108/F-01-2016-0001.

[43] S. Chapman, T. Kirks, J. Jost, Study on Interaction Modalities Between Humans and CPS in Sociotechnical Systems, in: T. Ahram, W. Karwowski, R. Taiar (Eds.), Human Systems Engineering and Design, Springer International Publishing, Cham, 2019: pp. 1044–1050.

[44] scikit-learn: machine learning in Python — scikit-learn 1.2.2 documentation, (n.d.). https://scikit-learn.org/stable/ (accessed May 25, 2023).

[45] TensorFlow Developers, TensorFlow (v2.12.0), (2023). https://doi.org/10.5281/zenodo.7764425.

[46] Linguistic Features · spaCy Usage Documentation, Linguistic Features. (n.d.). https://spacy.io/usage/linguistic-features#vectors-similarity (accessed June 6, 2023).

[47] D.B. Bisandu, R. Prasad, M.M. Liman, Data clustering using efficient similarity measures, Journal of Statistics and Management Systems. 22 (2019) 901–922. https://doi.org/10.1080/09720510.2019.1565443.

# APPENDIX A – SAMPLE SCHEDULE

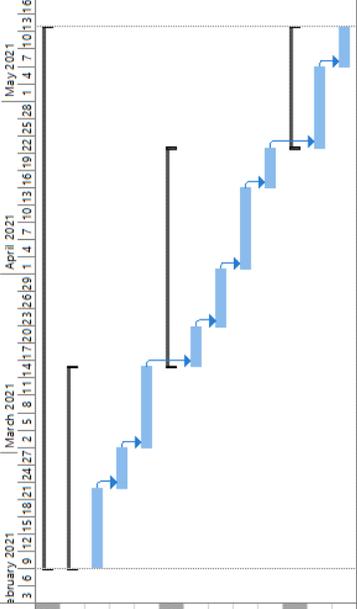

| ID | Task Mode | Task Name | Duration | Start | Finish | Task Status |
|---|---|---|---|---|---|---|
| 1 | | **CID Faculty Apartments** | **68 days** | **Tue 2/9/21** | **Thu 5/13/21** | |
| 2 | | **Structural framing** | **25 days** | **Tue 2/9/21** | **Mon 3/15/21** | Complete |
| 3 | | Columns | 10 days | Tue 2/9/21 | Mon 2/22/21 | Complete |
| 4 | | Beams | 5 days | Tue 2/23/21 | Mon 3/1/21 | Complete |
| 5 | | Floor deck | 10 days | Tue 3/2/21 | Mon 3/15/21 | Complete |
| 6 | | **Partition walls** | **28 days** | **Tue 3/16/21** | **Thu 4/22/21** | |
| 7 | | Wall tracks | 5 days | Tue 3/16/21 | Mon 3/22/21 | |
| 8 | | Wall studs | 8 days | Tue 3/23/21 | Thu 4/1/21 | |
| 9 | | Insulation | 10 days | Fri 4/2/21 | Thu 4/15/21 | |
| 10 | | Gypsum/Plaster boards | 5 days | Fri 4/16/21 | Thu 4/22/21 | |
| 11 | | **Finishing** | **15 days** | **Fri 4/23/21** | **Thu 5/13/21** | |
| 12 | | Painting | 10 days | Fri 4/23/21 | Thu 5/6/21 | |
| 13 | | Furniture | 5 days | Fri 5/7/21 | Thu 5/13/21 | |